\documentclass[runningheads]{llncs}
\usepackage{graphicx}
\begin{document}
\title{Robust Pancreatic Ductal Adenocarcinoma Segmentation with Multi-Institutional Multi-Phase  Partially-Annotated CT Scans}
\titlerunning{Robust Pancreatic Ductal Adenocarcinoma Segmentation}
%
%
%
\author{Ling Zhang\inst{1} \and
Yu Shi\inst{2} \and
Jiawen Yao\inst{1} \and
Yun Bian\inst{3} \and
Kai Cao\inst{3} \and 
Dakai Jin\inst{1} \and \\
Jing Xiao\inst{4} \and
Le Lu\inst{1}}
\authorrunning{L. Zhang et al.}
%
\institute{PAII Inc., Bethesda MD 20817, USA \and
Shengjing Hospital of China Medical University, Shenyang, China \and
Department of Radiology, Changhai Hospital, Shanghai 200433, China \and 
Ping An Technology Co., Ltd., Shenzhen, China \\
}
%
\maketitle              
\begin{abstract}
Accurate and automated tumor segmentation is highly desired since it has the great potential to increase the efficiency and reproducibility of computing more complete tumor measurements and imaging biomarkers, comparing to (often partial) human measurements. This is probably the only viable means to enable the large-scale clinical oncology patient studies that utilize medical imaging. Deep learning approaches have shown robust segmentation performances for certain types of tumors, e.g., brain tumors in MRI imaging, when a training dataset with plenty of pixel-level fully-annotated tumor images is available. However, more than often, we are facing the challenge that only (very) limited annotations are feasible to acquire, especially for hard tumors. Pancreatic ductal adenocarcinoma (PDAC) segmentation is one of the most challenging tumor segmentation tasks, yet critically important for clinical needs. 
Previous work on PDAC segmentation is limited to the moderate amounts of annotated patient images ($n$$<$300) from venous or venous+arterial phase CT scans. Based on a new self-learning framework, we propose to train the PDAC segmentation model using a much larger quantity of patients ($n$$\approx$1,000), with a mix of annotated and un-annotated venous or multi-phase CT images. Pseudo annotations are generated by combining two teacher models with different PDAC segmentation specialties on unannotated images, and can be further refined by a teaching assistant model that identifies associated vessels around the pancreas. A student model is trained on both manual and pseudo annotated multi-phase images. Experiment results show that our proposed method provides an absolute improvement of 6.3\% Dice score over the strong baseline of nnUNet trained on annotated images, achieving the performance (Dice = 0.71) similar to the inter-observer variability between radiologists.  

\end{abstract}
\section{Introduction}
For all machine learning-based tumor detection, characterization, and monitoring problems in cancer imaging, the volumetric segmentation of critical tumors plays an essential role \cite{bi2019artificial}. It serves the core for downstream processes of quantification, diagnosis, staging, prognosis, radiation planning, and treatment response prediction, all requiring a separate tumor segmentation step. In most of current clinical practices, tumor segmentation is still manually performed by clinicians, resulting in the consumption of labor, and low reproducibility of the derived biomarkers due to subjectivity. The tumor-related computerized models have been evaluated using relatively small-sized or moderate scale tumor datasets with expert annotations \cite{bi2019artificial,yamashita2020radiomic}. Deep learning methods can potentially increase the efficiency and reproducibility and make clinical/oncology patient studies scalable. For example, UNet \cite{ronneberger2015u} has recently been adopted for the objective and automated assessment of brain tumor treatment response in a multicentre study \cite{kickingereder2019automated}. However, it is usually infeasible and unconventional to construct such a large, well-organized, and volumetric-annotated tumor imaging dataset of 455 MRI scans by experts to train a fully-supervised deep model. We face a situation that the private and/or publicly existing full annotations are moderate to small-sized, while the unannotated imaging data can be huge. Furthermore, smaller-sized tumors are more difficult to segment, especially in CT scans with lower contrast than in MRI. Therefore fully automated and accurate (small) tumor segmentation in CT/MRI is still a challenging task to tackle.

Pancreatic ductal adenocarcinoma (PDAC), which constitutes 90\% of pancreatic cancers, is a dismal disease, with a 5-year overall survival rate only at 9\%. Less than 20\% of patients are eligible for the initial surgical resection. However, outcomes vary significantly even among the resected patients of the same TNM (tumor, node, and metastasis) stage receiving similar treatments. There is a critical and urgent need for additional predictive disease biomarkers to permit more personalized treatment. Radiological imaging provides the valuable noninvasive and informative information of the entire tumor. Subsequently, there are great interests in developing effective imaging-based biomarkers to stratify the group of resectable PDAC patients \cite{attiyeh2018survival} and predict gene mutation status from CT imaging \cite{attiyeh2019ct}, etc. Making these biomarkers to reach the clinical practices, a robust fully-automated PDAC segmentation model is desirable, as it can improve the objectiveness and enable the multicentre validation on a large-scale patient cohort. For the borderline resectable and locally-advanced PDACs, chemoradiation therapy is the suggested treatment option. One key step before each chemoradiation treatment is the manual segmentation and assessment of gross tumor volume, as a time-consuming and complex task requiring special expertise \cite{liang2020auto}.

Fully-automated segmentation of pancreatic tumor in CT scans is one of the most challenging tumor segmentation tasks, where previous state-of-the-art methods produce the Dice scores between 0.52 and 0.64 \cite{simpson2019large,isensee2018nnu,zhu2019multi,zhou2019hyper}, depending on the factors of tumor types (PDAC and others), utilized CT phases (arterial, venous), image quality, and consistency of annotations, and so on. Besides the complex abdominal structures, PDACs are quite variable in their shape, size, location, and enhancement patterns, demonstrating hypo-, iso-, or even hyper-enhancement. The heterogeneity of pancreas regions (i.e., pancreas tissue, duct, veins, and arteries) and the ill-defined tumor boundary make PDAC segmentation highly complicated even for radiologists. The inter-observer variability of this task is $\sim0.71$ in the Dice score, as reported in \cite{yamashita2020radiomic,liang2020auto}. 

In this paper, we propose a fully-automated and highly accurate PDAC segmentation method. Previous PDAC segmentation approaches \cite{isensee2018nnu,zhu2019multi,zhou2019hyper} are limited to the small to moderate amounts of annotated patients (patient number n$<$300) using venous or venous+arterial phase CT scans. In contrast, we train a model from a significantly larger patient population of n$\approx$1,000, including both (self-collected and publicly available) annotated and unannotated CT images covering multiple imaging phases, via the framework of self-learning \cite{zhang2018self,roth2019weakly,xie2020self}. Self-learning assumes that a deep model (student) trained from noisy annotations (teacher) has the potential to surpass the teacher \cite{zhang2018self}. Recent work finds new effective strategies to improve the student performance further, including adding regularization \cite{roth2019weakly} or noises \cite{xie2020self} to perturb the noisy annotations and generate noisy but informative annotations on a large unannotated external dataset \cite{xie2020self}. 
Specifically, (1) Our PDAC segmentation model is built upon a state-of-the-art medical image (particularly tumor) segmentation backbone nnUNet \cite{isensee2018nnu}, augmented using a new self-learning strategy that generates pseudo annotations on unannotated images by two teachers with different specialties, instead of traditionally by one teacher \cite{zhang2018self,roth2019weakly,xie2020self}. (2) We incorporate the semantic parsing of organs and vessels around the pancreas to further refine the pseudo annotations. (3) We demonstrate that our proposed method provides an absolute improvement of 6.3\% Dice score over the strong baseline of nnUNet \cite{isensee2018nnu} on multi-phase CT, achieving the highest fully-automated PDAC segmentation Dice score of 0.71 to date. Our results are substantially higher than the Dice scores of [0.52, 0.64] in \cite{simpson2019large,isensee2018nnu,zhu2019multi,zhou2019hyper} and comparable with the inter-observer variability \cite{yamashita2020radiomic,liang2020auto}. 

\section{Methods}

\begin{figure*}[!t]
   \begin{center}
   \begin{tabular}{c}
   \includegraphics[width=12cm]{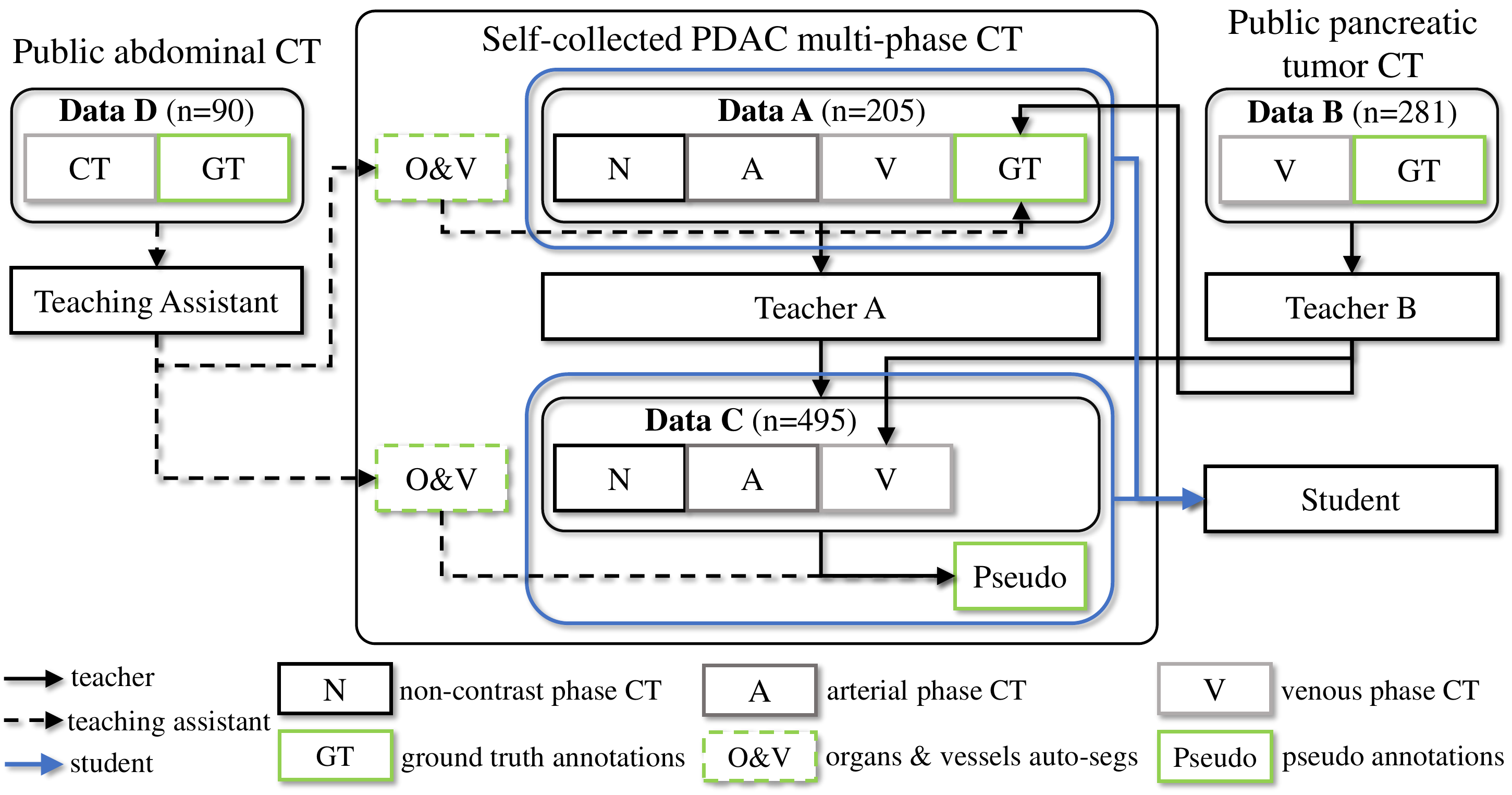}
   \end{tabular}
   \end{center}
   \caption
   { \label{fig1}
The proposed learning framework for pancreatic ductal adenocarcinoma (PDAC) segmentation on multi-phase CT scans. \textbf{Data A} and \textbf{Data C}: self-collected multi-phase CT datasets from two hospitals with and without PDAC annotations, respectively. \textbf{Data B}: a public venous phase CT dataset with pancreas and tumor annotations. \textbf{Data D}: public CT datasets with abdominal organ and vessel annotations. Pseudo annotations are generated by the teacher model A and B with different specialties. The student model is trained with the combined Data A and Data C and guided by the teaching assistant, which further corrects the pseudo pancreas annotations. 
} \vspace{-5mm} 
\end{figure*}

Fig. \ref{fig1} illustrates our proposed self-learning framework for PDAC segmentation using multi-phase CT scans. Given the self-collected Dataset A with only a moderate number of PDAC annotations available, our framework can effectively incorporate and utilize three other datasets (B, C, D) to significantly improve the PDAC segmentation performance, described as follows. 

In details, \textit{firstly}, we train a multi-phase pancreas and PDAC segmentation model on the private Dataset A with the registered non-contrast, arterial and venous phases CTs, denoted as Teacher A. \textit{Secondly}, Dataset B \cite{simpson2019large} is a public venous phase CT dataset including manual pixel-level annotations of pancreas and tumor and is used to train a segmentation model Teacher B. Note that since there is no pancreas annotations in our Dataset A, we use teacher B to segment the pancreas in the venous phase CT images in Dataset A to assist the training of teacher A\footnote[1]{We empirically find that a combined pancreas and tumor model segments PDAC tumors noticeably better than a single standalone tumor model.}. Teachers A and B have their specialized expertise depending on the different characteristics of Data A and B. \textit{Thirdly}, teacher A and teacher B respectively apply to segment the registered multi-phase CTs and the venous phase CTs in the self-collected Dataset C (a large scale but un-annotated multi-phase CT imaging study with PDACs). The resulted segmentation probability maps are adaptively combined to generate pseudo annotations of the pancreas and PDAC in Dataset C. \textit{Fourthly}, Dataset D \cite{gibson2018automatic} is a public abdominal CT dataset with annotations of the major abdominal organs \& vessels, and is employed here to train a Teaching Assistant to refine the pancreas annotations in Datasets A and C. \textit{Last}, we train a Student model on both Datasets A and C with manual and pseudo annotations (generated as above) of the pancreas and PDAC. For the multi-phase CT imaging registration as preprocessing, we use DEEDS \cite{heinrich2013towards}, which performs the best in a recent evaluation of abdominal CT imaging registration algorithms \cite{xu2016evaluation}.

{\bf nnUNet.} For the training of our segmentation models mentioned above, we use the nnUNet backbone \cite{isensee2018nnu} due to its high accuracy on several medical image segmentation tasks, such as abdominal organs, vessels, and tumors \cite{isensee2018nnu,kickingereder2019automated,isensee2019attempt}. The 3D UNet stage-1, which trains UNet on downsampled images, is used as the network architecture for a trade-off between the training efficiency and tumor segmentation accuracy. A combination of Dice and cross-entropy loss is utilized. We train the model to optimize the loss of both pancreas and PDAC. The model that produces the best Dice score of PDAC on the validation set is selected as the best PDAC segmentation model. In inference, we keep the non-background maximal connected component to remove false positives. For organ and vessel segmentation, we train a model to optimize the overall loss of all 17 classes \cite{gibson2018automatic}.  

{\bf Self-Learning.} Compared to previous self-learning methods of having one teacher, we introduce {\bf an additional teacher B to benefit from the knowledge} in the annotated public Dataset B \cite{simpson2019large}. While our self-collected Dataset A includes PDACs (mostly small size) located at the pancreas head and uncinate regions, Dataset B consists of a variety of size distributions of pancreatic tumors spanning over the whole pancreas. However, for Dataset B, (1) the pancreatic ducts, especially abnormal ones, can show similar appearances with PDACs in the venous phase CT; Other types of tumors in Dataset B, such as pancreatic neuroendocrine tumors, demonstrates different image enhancement patterns with PDACs. As a result, {\bf teacher B tends to identify some pancreatic ducts and normal pancreas tissues as tumors}. (2) On the other hand, the multi-phase (non-contrast, arterial, and venous) CT images in Dataset A can alleviate these difficulties by providing additional cues, such as the dynamic enhancement patterns of different structures. (3) By counting in all aspects, we generate the pseudo annotations for Dataset C by a weighted combining operator on the two teachers' segmentation probability maps. At the pancreas' head and uncinate, we assign a higher weight $\omega_{0}$ to teacher A (i.e., $1-\omega_{0}$ to teacher B), and at other regions, a higher weight $\omega_{1}$ to teacher B (i.e., $1-\omega_{1}$ to teacher A). As such, the teachers behave collaboratively like an ensemble when generating the pseudo annotations, which subsequently forces the student to learn from the reinforced ensemble model. (4) Moreover, the student's knowledge is expanded through learning on the large Dataset C that demonstrates more PDAC variations, allowing the student to learn beyond his teachers to be capable of segmenting more challenging images desirably. 

Note that we train the student model from scratch on the combination of Datasets A and C with manual and pseudo annotations. Such a strategy is found to be more effective than initializing the student with the teacher or first pre-training on un-annotated dataset and then finetuning on annotated dataset \cite{xie2020self}. Our overall self-learning training framework utilizing four datasets with different levels of annotations and patient distributions is shown in Fig. \ref{fig1}. In summary, the proposed teacher-student self-learning representation and training strategies enable our work and demonstrate the feasibility, as the largest study of this kind: utilizing three multi-institutional multi-phase PDAC CT imaging datasets.  

{\bf Organ and Vessel Parsing.} The semantic image parsing (of organs and vessels) component acts as a teaching assistant. It corrects and refines the pseudo annotations of the pancreas in Datasets A and C by masking out the (previously produced) pancreas annotations that belong to different vessel classes (i.e., portal and splenic vein, superior mesenteric vein and artery, and truncus coeliacus). As such, the final student model of pancreas+PDAC segmentation is encouraged to be more focused on learning the pancreas region (with reinforced and corrected pseudo annotation) and distinguishing Pancreas and PDAC from vessels. We observe that without the teaching assistant, some vessel regions around the pancreas tend to be segmented as the pancreas or PDAC even after self-learning.

\section{Experimental Results}

{\bf Datasets.} Four multi-institutional datasets (total n=1,071) are used in this work. \textbf{Dataset A}, including 205 patients with PDACs (the mean tumor size is 2.5 cm), is collected from hospital A with non-contrast, pancreatic (late arterial), and venous phases of CT scans. Such a multi-phase CT imaging setting is the standardized protocol for depiction, staging, and resectability evaluations of PDAC, specified in the National Comprehensive Cancer Network (NCCN) guidelines \cite{nccnpdac}. The median imaging spacing is 0.70$\times$0.70$\times$3mm in [X,Y,Z]. PDAC tumors are manually traced and annotated on the pancreatic phase slice by slice by one radiologist with 18 years of experience in pancreatic imaging. \textbf{Dataset B}, including 281 patients with pancreatic tumor annotations, is a public dataset provided by Memorial Sloan Kettering Cancer Center \cite{simpson2019large}. The median imaging spacing is 0.80$\times$0.80$\times$2.5mm. \textbf{Dataset C}, including 495 patients with PDACs (no manual annotations), is collected from hospital C with non-contrast, early arterial, and venous phase CTs used in this work. The median imaging spacing is 0.68$\times$0.68$\times$3mm. \textbf{Dataset D} is a combination of two public datasets (described in \cite{gibson2018automatic}), including 90 patients' abdominal CTs with annotated organs and vessels up to 14 classes. An engineer manually annotates three additional vessel classes (superior mesenteric vein and artery, and truncus coeliacus) in 46 CTs and completes some CTs without annotations of the portal and splenic veins under the supervision of a board-certified radiologist. An initial nnUNet network is trained to segment the remaining CTs; the segmented pseudo masks of the four classes of vessels are manually corrected by the engineer. Finally, the second nnUNet (as teaching assistant) is trained on all 90 CTs with 17 classes of annotations. 

{\bf Implementation details.} For the nnUNet~\cite{isensee2018nnu} training, most parameters are set by default. We find that increasing the batch size or changing the optimizer produces similar performance. Multi-phase CT images are directly concatenated as input channels to feed the network. The input-level fusion is widely adopted in the multi-modality tumor segmentation tasks \cite{zhou2019review}. We compare the input-level fusion with a layer-level fusion, i.e., the hyper-pairing network \cite{zhou2019hyper}, which previously shows the state-of-the-art PDAC segmentation performance for multi-phase CTs. The implementations remain the same with the nnUNet framework except that the backbone network is replaced with the hyper-pairing UNet. Only arterial and venous phases are used as in \cite{zhou2019hyper}, and our GPU cannot feed in the hyper-pairing UNet with three CT phases. The phase augmentation and pairing loss in \cite{zhou2019hyper} are not implemented. The weights for combining the two teachers' segmentation probability maps are set as $\omega_{0}=0.8$ and $\omega_{1}=0.6$, with the first 60\% pancreas volume (starting from the left-most slice occupied by the pancreas annotations in the sagittal direction) being roughly treated as the head and uncinate; and the remaining 40\% as other locations. The teacher models are trained with 200 epochs (250 batches per epoch), and the student models with 33 epochs (1500 batches per epoch). Each training process takes one day on a NVIDIA Titan RTX-6000 GPU. For training the teacher and teaching assistant models, a random training-validation splitting is used. For comparing the models' performance, five-fold cross-validation is used. The results on Dataset A are reported, and Dice coefficient is used as the evaluation metric.


{\bf Quantitative Results \& Discussion.} Table \ref{results} shows the PDAC segmentation results on Dataset A by different methods. The nnUNet trained on a single venous phase CT has a mean Dice score of 0.522, comparable with the performances reported in \cite{isensee2018nnu,zhou2019hyper} when evaluating UNet only on the venous phase CT. A single pancreatic (late arterial) phase substantially improves the performance to 0.630. This is consistent with the clinical guideline \cite{nccnpdac} that PDACs are best visible in the pancreatic phase. Non-contrast phase has the lowest performance. But when it is combined with the other two phases, the PDAC segmentation performance can be further increased to 0.646, surpassing the previous reported highest Dice score of 0.639 \cite{zhou2019hyper}. This is likely because the heterogeneity of the pancreas regions (e.g., tissue, pancreas duct, vessels) showing in the contrast-enhanced CT phases becomes more homogeneous in the non-contrast phase. The hyper-pairing \cite{zhou2019hyper} nnUNet does not improve the accuracy in our experiment. However, we acknowledge that our implementation may not be fully delicately optimized, e.g., it may need to be trained for more epochs, given that the hyper-pairing network is twice larger than a single network.

\begin{table}[t]
\caption{PDAC segmentation results on Data A by 5-fold cross-validation. TA: teaching assistant model trained on Data D. Results are reported as mean$\pm$std. $^{*}$ is an average value of the two inter-observer variability values in \cite{yamashita2020radiomic} which are median Dice scores. As a comparison, our final student model has a median Dice of 0.755.}
\label{results}
\footnotesize
\centering
\begin{tabular}{l|l|l}
\hline
Methods: CT phases                           & Training Data           & Dice          \\ \hline
nnUNet: venous (Teacher B)                               & Data B & 0.395$\pm$0.290               \\ \hline
nnUNet: non-contrast                          & Data A & 0.476$\pm$0.221         \\
nnUNet: pancreatic                             & Data A & 0.630$\pm$0.230         \\
nnUNet: venous                               & Data A & 0.522$\pm$0.250         \\
nnUNet: pancreatic+venous                      & Data A & 0.629$\pm$0.222         \\
Hyper-pairing \cite{zhou2019hyper} nnUNet: pancreatic+venous        & Data A & 0.604$\pm$0.230         \\
nnUNet: 3-phases (Teacher A)         & Data A & 0.646$\pm$0.199        \\ \hline
nnUNet+Self-learn: 3-phases (Student)      & Data A, B, C & 0.698$\pm$0.174       \\
nnUNet+Self-learn+TA: 3-phases (Student) & Data A, B, C, D & \textbf{0.709$\pm$0.159}         \\ \hline
Inter-observer (CT \cite{yamashita2020radiomic} /MRI \cite{liang2020auto})      & & 0.697$^{*}$ / 0.710 \\ \hline
\end{tabular}\vspace{-5mm}
\end{table}

   \begin{figure*}[t]
   \begin{center}
   \begin{tabular}{c}
   \includegraphics[width=12.2cm]{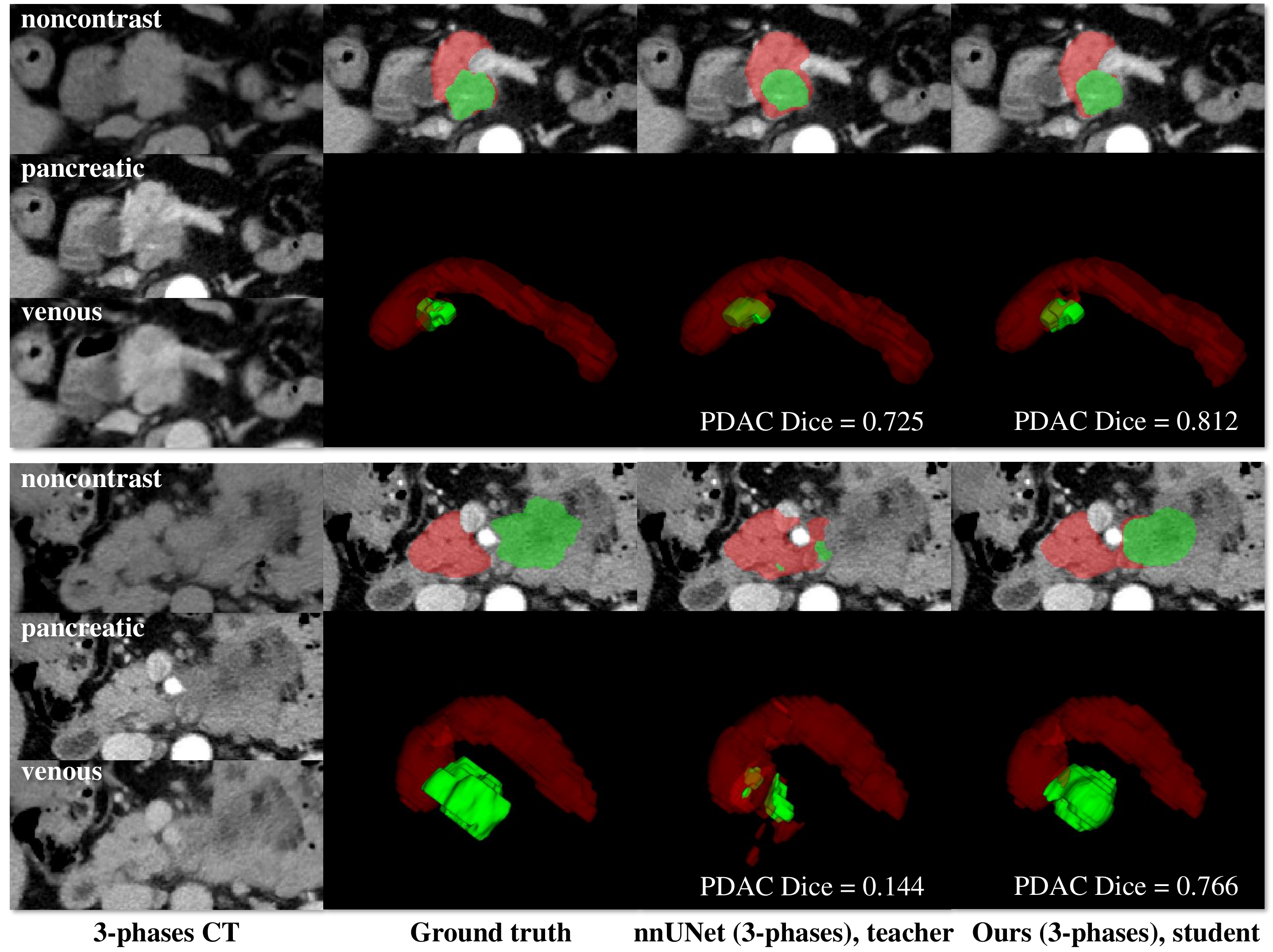}
   \end{tabular}
   \end{center}
   \caption
   { \label{fig2}
Two qualitative examples of Pancreas (red) and PDAC (green) segmentation on multi-phase CT images. Upper panel: The student model further improves the teacher's accuracy by identifying/delineating more accurate PDAC boundaries. Lower panel: The student model substantially outperforms its teacher in segmenting a large tumor with irregular shape located at the pancreas uncinate. In both cases, the student model successfully identifies vessels around the pancreas (i.e., portal and splenic veins, superior mesenteric vein and artery) by excluding them from segmentation.
}
\end{figure*}
More importantly, compared to the above strong baselines of nnUNet trained on three phases, our proposed self-learning approach substantially improves the PDAC segmentation performance in Dice scores from 0.646 to 0.698. By training with the pseudo annotations further corrected by the teaching assistant, the student model reaches a Dice score of 0.709 (roughly $10.5\%$ absolute Dice improvement than the previous state-of-the-art work \cite{zhou2019hyper}). This also achieves a similar performance compared to the inter-observer variability between radiologists \cite{yamashita2020radiomic,liang2020auto}. The learned student model can improve the teacher's performance in general and is especially better in some cases, which are difficult for the teacher, e.g., large tumors with irregular shape, veins. Two illustrative pancreas and PDCA segmentation examples are shown in Fig. \ref{fig2}.

\section{Conclusion}
Fully automated and accurate segmentation of pancreatic ductal adenocarcinoma (PDAC) is one of the most challenging tumor segmentation tasks, in the aspects of complex abdominal structures, large variations in morphology and appearance, low image contrast and fuzzy/uncertain boundary, etc. Previous studies introduce the cascade UNet for segmenting venous phase CT and hyper-pairing network for segmenting venous+arterial phases CT and achieving mean Dice scores of 0.52 and 0.64, respectively. By incorporating nnUNet into a new self-learning framework with two teachers and one teaching assistant to segment three-phases of CT scans, our method reaches a Dice coefficient of 0.71, similar to the inter-observer variability between radiologists. This provides promise that a radiologist-level performance for accurate PDAC tumor segmentation in multi-phase CT imaging can be achieved through our computerized method.


\bibliographystyle{splncs04}
\bibliography{ref}

\begin{thebibliography}{10}
\providecommand{\url}[1]{\texttt{#1}}
\providecommand{\urlprefix}{URL }
\providecommand{\doi}[1]{https://doi.org/#1}

\bibitem{nccnpdac}
{NCCN Clinical Practice Guidelines in Oncology (NCCN Guidelines®) Pancreatic
  Adenocarcinoma}.
  \url{https://www2.tri-kobe.org/nccn/guideline/pancreas/english/pancreatic.pdf}

\bibitem{attiyeh2018survival}
Attiyeh, M.A., Chakraborty, J., Doussot, A., Langdon-Embry, L., Mainarich, S.,
  G{\"o}nen, M., Balachandran, V.P., D’Angelica, M.I., DeMatteo, R.P.,
  Jarnagin, W.R., et~al.: Survival prediction in pancreatic ductal
  adenocarcinoma by quantitative computed tomography image analysis. Ann. Surg.
  Oncol.  \textbf{25}(4),  1034--1042 (2018)

\bibitem{attiyeh2019ct}
Attiyeh, M.A., Chakraborty, J., McIntyre, C.A., Kappagantula, R., Chou, Y.,
  Askan, G., Seier, K., Gonen, M., Basturk, O., Balachandran, V.P., et~al.:
  {CT} radiomics associations with genotype and stromal content in pancreatic
  ductal adenocarcinoma. Abdom. Radiol.  \textbf{44}(9),  3148--3157 (2019)

\bibitem{bi2019artificial}
Bi, W.L., Hosny, A., Schabath, M.B., Giger, M.L., Birkbak, N.J., Mehrtash, A.,
  Allison, T., Arnaout, O., Abbosh, C., Dunn, I.F., et~al.: Artificial
  intelligence in cancer imaging: clinical challenges and applications. CA:
  Cancer J. Clin.  \textbf{69}(2),  127--157 (2019)

\bibitem{gibson2018automatic}
Gibson, E., Giganti, F., Hu, Y., Bonmati, E., Bandula, S., Gurusamy, K.,
  Davidson, B., Pereira, S.P., Clarkson, M.J., Barratt, D.C.: Automatic
  multi-organ segmentation on abdominal {CT} with dense v-networks. TMI
  \textbf{37}(8),  1822--1834 (2018)

\bibitem{heinrich2013towards}
Heinrich, M.P., Jenkinson, M., Papie{\.{z}}, B.W., Brady, S.M., Schnabel, J.A.:
  Towards realtime multimodal fusion for image-guided interventions using
  self-similarities. In: Mori, K., Sakuma, I., Sato, Y., Barillot, C., Navab,
  N. (eds.) Medical Image Computing and Computer-Assisted Intervention --
  MICCAI 2013. pp. 187--194. Springer Berlin Heidelberg, Berlin, Heidelberg
  (2013)

\bibitem{isensee2019attempt}
Isensee, F., Maier-Hein, K.H.: {An attempt at beating the 3D U-Net}. arXiv
  preprint arXiv:1908.02182  (2019)

\bibitem{isensee2018nnu}
Isensee, F., Petersen, J., Klein, A., Zimmerer, D., Jaeger, P.F., Kohl, S.,
  Wasserthal, J., Koehler, G., Norajitra, T., Wirkert, S., et~al.: nn{U}-{N}et:
  Self-adapting framework for u-net-based medical image segmentation. arXiv
  preprint arXiv:1809.10486  (2018)

\bibitem{kickingereder2019automated}
Kickingereder, P., Isensee, F., Tursunova, I., Petersen, J., Neuberger, U.,
  Bonekamp, D., Brugnara, G., Schell, M., Kessler, T., Foltyn, M., et~al.:
  Automated quantitative tumour response assessment of {MRI} in neuro-oncology
  with artificial neural networks: a multicentre, retrospective study. Lancet
  Oncol.  \textbf{20}(5),  728--740 (2019)

\bibitem{liang2020auto}
Liang, Y., Schott, D., Zhang, Y., Wang, Z., Nasief, H., Paulson, E., Hall, W.,
  Knechtges, P., Erickson, B., Li, X.A.: Auto-segmentation of pancreatic tumor
  in multi-parametric {MRI} using deep convolutional neural networks.
  Radiother. Oncol.  \textbf{145},  193--200 (2020)

\bibitem{ronneberger2015u}
Ronneberger, O., Fischer, P., Brox, T.: U-net: Convolutional networks for
  biomedical image segmentation. In: Navab, N., Hornegger, J., Wells, W.M.,
  Frangi, A.F. (eds.) Medical Image Computing and Computer-Assisted
  Intervention -- MICCAI 2015. pp. 234--241. Springer International Publishing,
  Cham (2015)

\bibitem{roth2019weakly}
Roth, H., Zhang, L., Yang, D., Milletari, F., Xu, Z., Wang, X., Xu, D.: Weakly
  supervised segmentation from extreme points. In: Zhou, L., Heller, N., Shi,
  Y., Xiao, Y., Sznitman, R., Cheplygina, V., Mateus, D., Trucco, E., Hu, X.S.,
  Chen, D., Chabanas, M., Rivaz, H., Reinertsen, I. (eds.) Large-Scale
  Annotation of Biomedical Data and Expert Label Synthesis and Hardware Aware
  Learning for Medical Imaging and Computer Assisted Intervention. pp. 42--50.
  Springer International Publishing, Cham (2019)

\bibitem{simpson2019large}
Simpson, A.L., Antonelli, M., Bakas, S., Bilello, M., Farahani, K.,
  Van~Ginneken, B., Kopp-Schneider, A., Landman, B.A., Litjens, G., Menze, B.,
  et~al.: A large annotated medical image dataset for the development and
  evaluation of segmentation algorithms. arXiv preprint arXiv:1902.09063
  (2019)

\bibitem{xie2020self}
Xie, Q., Luong, M.T., Hovy, E., Le, Q.V.: Self-training with noisy student
  improves imagenet classification. In: Proceedings of the IEEE/CVF Conference
  on Computer Vision and Pattern Recognition. pp. 10687--10698 (2020)

\bibitem{xu2016evaluation}
Xu, Z., Lee, C.P., Heinrich, M.P., Modat, M., Rueckert, D., Ourselin, S.,
  Abramson, R.G., Landman, B.A.: Evaluation of six registration methods for the
  human abdomen on clinically acquired {CT}. IEEE Trans. Biomed. Eng.
  \textbf{63}(8),  1563--1572 (2016)

\bibitem{yamashita2020radiomic}
Yamashita, R., Perrin, T., Chakraborty, J., Chou, J.F., Horvat, N., Koszalka,
  M.A., Midya, A., Gonen, M., Allen, P., Jarnagin, W.R., et~al.: Radiomic
  feature reproducibility in contrast-enhanced {CT} of the pancreas is affected
  by variabilities in scan parameters and manual segmentation. Eur. Radiol.
  \textbf{30}(1),  195--205 (2020)

\bibitem{zhang2018self}
Zhang, L., Gopalakrishnan, V., Lu, L., Summers, R.M., Moss, J., Yao, J.:
  Self-learning to detect and segment cysts in lung {CT} images without manual
  annotation. In: ISBI 2018. pp. 1100--1103 (2018)

\bibitem{zhou2019review}
Zhou, T., Ruan, S., Canu, S.: A review: Deep learning for medical image
  segmentation using multi-modality fusion. Array p. 100004 (2019)

\bibitem{zhou2019hyper}
Zhou, Y., Li, Y., Zhang, Z., Wang, Y., Wang, A., Fishman, E.K., Yuille, A.L.,
  Park, S.: Hyper-pairing network for multi-phase pancreatic ductal
  adenocarcinoma segmentation. In: Shen, D., Liu, T., Peters, T.M., Staib,
  L.H., Essert, C., Zhou, S., Yap, P.T., Khan, A. (eds.) Medical Image
  Computing and Computer Assisted Intervention -- MICCAI 2019. pp. 155--163.
  Springer International Publishing, Cham (2019)

\bibitem{zhu2019multi}
Zhu, Z., Xia, Y., Xie, L., Fishman, E.K., Yuille, A.L.: Multi-scale
  coarse-to-fine segmentation for screening pancreatic ductal adenocarcinoma.
  In: Shen, D., Liu, T., Peters, T.M., Staib, L.H., Essert, C., Zhou, S., Yap,
  P.T., Khan, A. (eds.) Medical Image Computing and Computer Assisted
  Intervention -- MICCAI 2019. pp. 3--12. Springer International Publishing,
  Cham (2019)

\end{thebibliography}
\end{document}